\documentclass[useAMS,usenatbib]{mn2e}
\usepackage{lineno,color}
\RequirePackage{lineno}
\usepackage{amssymb}
\usepackage{amsmath}
\usepackage{txfonts}
\usepackage{graphicx}
\usepackage{natbib}
\usepackage{rotating}
\usepackage{url}
\usepackage{epsfig}
\usepackage{xspace}
\usepackage{array}
\usepackage{multirow}
 \usepackage[T1]{fontenc} 
 \usepackage{aecompl}

\def\ltsima{$\; \buildrel < \over \sim \;$}
\def\simlt{\lower.5ex\hbox{\ltsima}} 
\def\gtsima{$\; \buildrel > \over \sim \;$}
\def\simgt{\lower.5ex\hbox{\gtsima}} 

\def\deg{\hbox{$^\circ$}}
\def\ltsima{$\; \buildrel < \over \sim \;$}
\def\simlt{\lower.5ex\hbox{\ltsima}} 
\def\gtsima{$\; \buildrel > \over \sim \;$}
\def\simgt{\lower.5ex\hbox{\gtsima}} 

\def\deg{\hbox{$^\circ$}}
\def\gr{$\gamma$-ray}
\def\g{$\gamma$}

\def\bi {\begin{itemize}}
\def\ei {\end{itemize}}

\def\xmm{\textit {XMM-Newton}}
\def\szk{\textit {Suzaku}}
\def\nst{\textit {NuSTAR}}

\def\swift{\textit {Swift}}

\def\psrb{PSR~B1259$-$63}

\def\rxp {2RXP~J130159.6$-$635806}
\def\deg {$^\circ$}

\def\F{{\em Fermi}}

\def\L{LAT}

\def\psrb{PSR~B1259$-$63}

\begin{document}
\title[Multi-wavelength Observations of  \psrb\ around the 2014 Periastron Passage]{Multi-wavelength Observations of  the Binary System \psrb/LS~2883 around the 2014 Periastron Passage} 
\author[M. Chernyakova et.al.]{M. Chernyakova$^{1,2}$,  A. Neronov $^{3}$, B. van Soelen $^{4}$, P. Callanan  $^{5}$,  L. O'Shaughnessy$^{5}$,  \newauthor Iu. Babyk$^{1,2,18}$, S. Tsygankov$^{6}$, Ie. Vovk$^{7}$, R. Krivonos $^{8,16}$, J. A. Tomsick $^{8}$, D. Malyshev$^{3}$, \newauthor J. Li $^{9}$, K. Wood $^{10}$, D. Torres $^{9}$, S. Zhang $^{11}$, P. Kretschmar $^{12}$, M.V. McSwain $^{13}$, \newauthor  D. Buckley $^{14,17}$, C. Koen $^{15}$ \\ 
$^{1}$ Dublin City University, Dublin 9, Ireland\\
$^{2}$ Dublin Institute for Advanced Studies, 31 Fitzwilliam Place, Dublin 2, Ireland\\
$^{3}$ INTEGRAL Science Data Center, Chemin d'\'Ecogia 16, 1290 Versoix, Switzerland\\
$^{4}$ Department of Physics, University of the Free State, P.O. Box 339 Bloemfontein, Free State, 9300, Republic of South Africa \\
$^{5}$ Department of Physics, University College Cork, Cork, Ireland \\
$^{6}$ Tuorla Observatory, Department of Physics and Astronomy, University of Turku, Vaisalantie 20, FI-21500 Piikkio, Finland\\
$^{7}$ Max Planck Institut für Physik, Föhringer ring 6, 80805, Munich, Germany \\
$^{8}$ Space Sciences Laboratory, 7 Gauss Way, University of California, Berkeley, CA 94720-7450, USA\\
$^{9}$ Institute of Space Sciences (CSIC-IEEC), Campus UAB, Carrer de Can Magrans, s/n 08193, Barcelona, Spain \\
$^{10}$ Space Science Division, Naval Research Laboratory, USA\\
$^{11}$ Laboratory for Particle Astrophysics, Institute of High Energy Physics, Beijing 100049, China\\
$^{12}$ European Space Astronomy Centre (ESA/ESAC), Science Operations Department, Villanueva de la Canada (Madrid), Spain \\
$^{13}$ Department of Physics, Lehigh University, 16 Memorial Drive East, Bethlehem, PA, USA \\
$^{14}$ South African Astronomical Observatory, Observatory Road, Observatory 7935, South Africa \\
$^{15}$ Department of Statistics, University of the Western Cape, Bellville 7535, South Africa \\
$^{16}$ Space Research Institute, Russian Academy of Sciences, Profsoyuznaya 84/32, 117997 Moscow, Russia \\
$^{17}$ South African Large Telescope, PO Box 9, Observatory 7935, South Africa\\
$^{18}$ Main Astronomical Observatory of National Academy of Science of Ukraine, Academica Zabolotnogo str., 27, 03680, Kyiv, Ukraine\\
$^{19}$ Instituci\'o Catalana de Recerca i Estudis Avan\c{c}ats (ICREA) Barcelona, Spain
}

\date{Received $<$date$>$  ; in original form  $<$date$>$ }
\pagerange{\pageref{firstpage}--\pageref{lastpage}} \pubyear{2014}

\maketitle
\label{firstpage}
\begin{abstract}
{ We report on the results of the extensive multi-wavelength campaign from optical to GeV \g-rays of the 2014 periastron passage of \psrb, which is a unique high-mass \g-ray emitting binary system with a young pulsar companion. 
Observations demonstrate the stable nature of the post-periastron GeV flare and prove the coincidence of the flare with the start of rapid decay of the H$\alpha$ equivalent width, usually interpreted as a disruption of the Be stellar disk. Intensive X-ray observations reveal changes in the X-ray spectral behaviour happening at the moment of the GeV flare. We demonstrate that these changes can be naturally explained as a result of synchrotron cooling of monoenergetic relativistic electrons injected into the system during the GeV flare.   
}
\end{abstract}

\section{Introduction} \label{section-intro}

In the binary system \psrb\ a 47.76 ms radio pulsar is in a highly
eccentric orbit ($e \approx 0.87, P \approx 3.4$ years) around the massive O9.5Ve star LS~2883 \citep{Johnston1992,distance}. The optical spectrum of the companion  shows 
evidence of an equatorial disk (thus this star is generally
classified as a Be star), which is thought to be inclined with respect to the orbital plane \citep{melatos1995}. The pulsar crosses the disk plane twice each orbit, just before and  after the
periastron passage. The minimum approach between the pulsar and
massive star is about $\sim 0.9$ astronomical unit \citep{distance}, which
is roughly the size of the equatorial disk \citep{Johnston1992}. 
 Interaction between the relativistic pulsar wind and the wind and photon field of the Be star is believed to give rise to the observed unpulsed emission. X-ray emission is observed throughout the orbit \citep[e.g.][]{Hirayama1999, Chernyakova2006} and the unpulsed radio, GeV and TeV emission is observed within a few months of periastron passage \citep{Johnston1999,Johnston2005,kirk99,Abdo2011_b1259,Chernyakova_psrb11, Aharonian_B1259_2004_pass,psrb1259_hess09}. 
 
 The presence of a nearby X-ray pulsar, \rxp, located only 10 arcminutes away from PSR B1259-63 makes it difficult to observe the source with a non-imaging instrument, like \szk\ HXD-PIN detectors, or instruments with a  low angular resolution, like INTEGRAL. For these instruments both sources influence the observed emission, and in order to reconstruct the spectral shape of \psrb\ it is necessary
to know the spectrum of \rxp . Recent  \textit{NuSTAR} observations of the \rxp\ \citep{krivonos15} allowed us to reconstruct the broad band,   0.5 - 60 keV X-ray spectrum of \psrb\, as observed by \szk\ in 2011, and  
 by INTEGRAL  in 2015. 

The most puzzling feature of the orbital modulation of the source is the huge flare in the GeV band starting ten days after the post-periastron passage of the pulsar through the stellar disk with no obvious counterpart at any other wavelength \citep{Abdo2011_b1259,Chernyakova_psrb11}. The origin of this flare was widely discussed in the literature (see e.g. \citet{Abdo2011_b1259,Petri11,khan12,Kirk13,Dubus13}), but the lack of observational data prevented any firm conclusions being made.
It was even not clear in advance whether the GeV flare  would repeat in 2014.

In this paper we present the results of the extensive multiwavelength campaign of the latest PSR B1259-63 periastron passage, which happened on 2014 May 4 (t$_p$=MJD56781.42). One of the main aims of this campaign was multiwavelength monitoring of \psrb\ around the periastron and the moment of GeV flare in order to reveal the true nature of the flare. The paper is organised the following way. In section \ref{sec:Opt} we present results of optical monitoring with  the Southern African Large Telescope (SALT) and SAA0 1.9-m telescopes. In section \ref{xr} details of X-ray data analysis along with the results of X-ray monitoring with \textit{SWIFT}, \textit{NuSTAR}, and INTEGRAL are given. In this section we also present for the first time the
broadband X-ray spectrum of \psrb\ as seen in 2010 by \szk . Section \ref{fermi} explains details of \F-\L\ data analysis. Finally all the obtained results are discussed in  section \ref{disc}.


\section{Optical Spectroscopy } \label{sec:Opt}

\begin{figure*}
\resizebox{0.9\hsize}{!}{\includegraphics[angle=0]{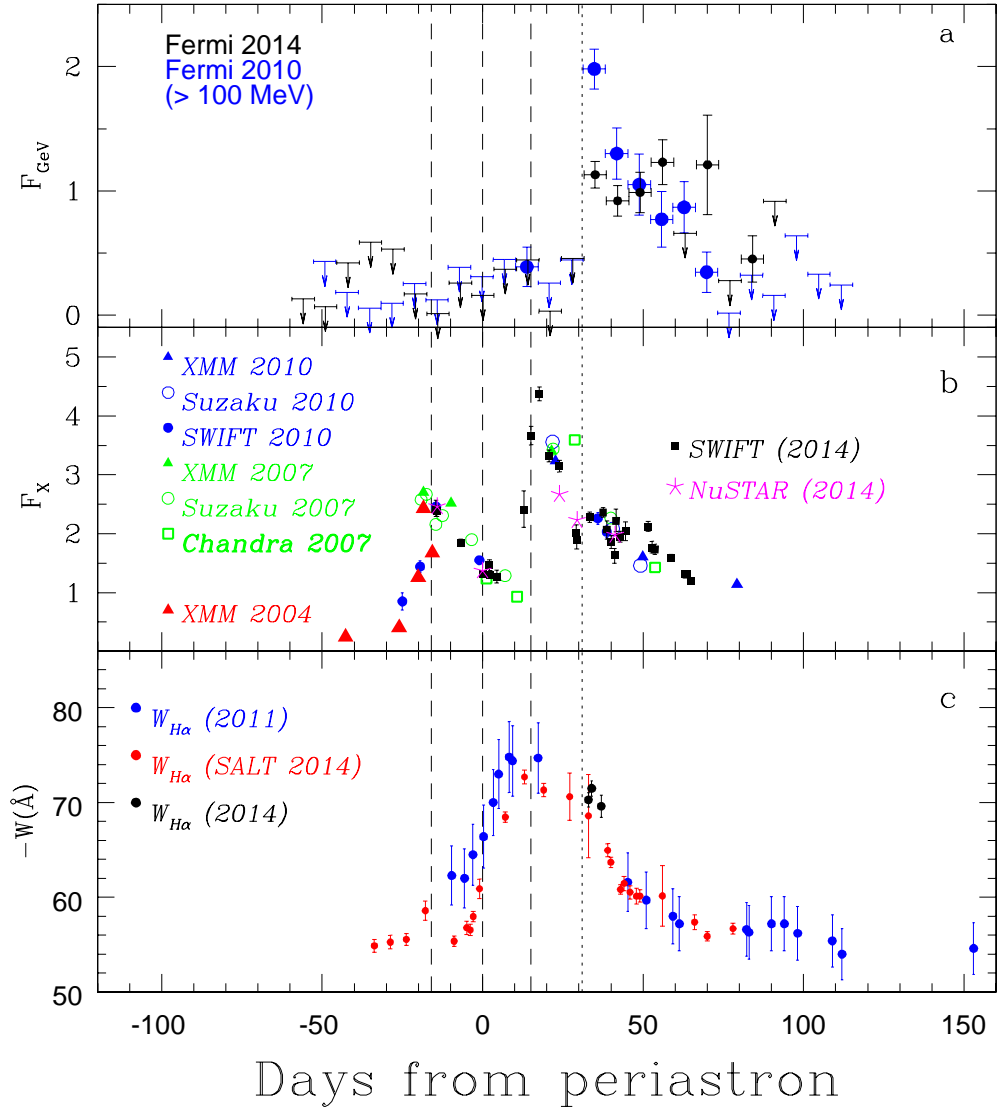}}
\caption{Orbital light curves of \psrb\ around periastron for several
  passages. Dashed lines correspond to the periastron and to the moments of disappearance (first non-detection) and reappearance (first detection) of the pulsed emission, as observed in 2010 \citep{Abdo2011_b1259}. Dotted line corresponds to the first appearance of the detection in GeV band at a day time scale. \textit{ Panel a}:  \F-\L\ flux measurements in
  the $E > 100$~MeV energy range for the 2010 and 2014 periastron passages with a weekly bin size. Flux is given in $10^{-6}$~cm$^{-2}$~s$^{-1}$. \textit{Panel b}:
  X-ray fluxes from 2014 (this work) and three last periastron passages \citep{Abdo2011_b1259,
    Chernyakova2009}. Flux is given in $10^{-11}$~erg~cm$^{-2}$~s$^{-1}$. The typical error of the X-ray data is smaller than the size of the symbols.    \textit{Panel c}: Evolution of the equivalent widths of H$\alpha$. 
}

\label{fig:MW_LC} 
\end{figure*}

\subsection{SAA0 1.9-m telescope \label{saao19}}
Spectroscopic observations of LS 2883, the primary of the system, were performed with the SAA0 1.9-m telescope between the 2014 June 6 and 2014 June 11. The Cassegrain spectrograph was set with grating number 5 to achieve a wavelength coverage of 6350-7050 \AA\, with a blaze of 6800 \AA\ and a resolving power R = $\lambda/\Delta \lambda$ = 6563 in the vicinity of the H$_\alpha$ line. 

LS 2883 was observed for a total of 3 nights using 30s exposure times. 240 individual spectra were taken on the first night, 245 on the second and 196 on the third. 
Copper-Neon comparison lamp spectra were obtained throughout each night for wavelength calibration. 

The spectra were flat fielded, extracted and wavelength calibrated using the standard slit spectroscopy routines in IRAF. To improve the signal-to-noise (S/N) of weak lines the spectra for each night were coadded. The mean spectra were then normalised to a unit continuum. 

The equivalent widths of the H$\alpha$  and He I(6678 \AA) lines were
 measured by integrating over the emission line profiles. Due to the high S/N of the co-added spectra, the errors in W(H$\alpha$) and W(He I(6678 \AA)) are due principally to the placement of the continuum. The larger percentage errors in W(He I(6678 \AA)) is a result of an absorption feature in the vicinity, and the lower line strength. By choosing appropriate upper and lower bounds on the continuum levels, the error margins on the equivalent widths were calculated.

Further observations of LS 2883 were made between the UT dates 2014 June 11 and 2014 June 17. The Cassegrain spectrograph was set with grating number 6 to achieve a wavelength coverage of 4100-5400 \AA\, with a blaze of 4600 \AA\ and a resolving power R = $\lambda/\Delta \lambda$ = 2430 in the vicinity of the H$_\beta$ line.


In this mode LS 2883 was observed for 5 more nights with either 1 or 2 spectra taken each night. The exposure times each night varied from 350s to 700s depending on the number of spectra taken. Copper-argon comparison lamp spectra were obtained before and after each sequence for wavelength calibration. 



An attempt was made to infer the equivalent width of the H$_\alpha$ line using the equivalent width of the H$_\beta$ line. Simultaneous measurements of the H$_\alpha$ and H$_\beta$ lines have been previously made and the ratio of W(H$_\alpha$) to W(H$_\beta$) was found to be $11.74 \pm 0.67$ \citep{distance}.  The application of this result to our H$_\beta$ data  allowed for the deduction of W(H$_\alpha$) from the measured W(H$_\beta$) values. The larger error margins on the converted H$_\alpha$ data relative to the H$_\alpha$ data from the first three nights of observations is due to a combination of larger percentage errors on the W(H$_\beta$) values and the propagation of the uncertainty in the W(H$_\alpha$)/W(H$_\beta$) relation from \cite{distance}.

The data of \cite{distance} were taken at a phase of ~0.62, while our observations are made at a phase of ~0.02. The assumption that the W(H$_\alpha$)/W(H$_\beta$) ratio remains constant with phase is not completely justified considering that the physical conditions of the disk drastically varies as the pulsar moves through. However, spectroscopic observations from SALT were made simultaneously for two of the nights. The agreement between the two data sets suggests that the W(H$_\alpha$)/W(H$_\beta$) relation does not change with phase by more than the measured uncertainty in the equivalent widths of \cite{distance}.


\begin{table}
\begin{center}
\caption{Journal of SAA0 1.9-m telescope 2014 observations of the source PSR B1259-63 around the time of GeV flare.}\label{tab:1.9-m} 
\begin{tabular}{|c|c|c|c|c|c|}
\hline 
\multirow{1}{*}{Date} & \multicolumn{1}{c}{MJD} & $t-t_{p}$ &$\varphi$& $-W_{H\alpha}$  & $-W_{6678}$ \tabularnewline
 & (days) & (days) & &(\r{A}) & (\r{A}) \tabularnewline
\hline 
2014-06-06 & 56814.74 & 33.3 & 294.8& 70.28 $\pm$ 0.73 & 0.44 $\pm$ 0.05\tabularnewline
2014-06-07 & 56815.77 & 34.3 & 295.8& 71.48 $\pm$ 0.81 & 0.43 $\pm$ 0.05\tabularnewline
2014-06-10 & 56818.71 & 37.3 & 298.4& 69.61 $\pm$ 1.16 & 0.44 $\pm$ 0.05\tabularnewline
\hline 
2014-06-11 & 56819.87 & 38.4 &  299.3&71.17 $\pm$ 4.39 &  -\tabularnewline
2014-06-12 & 56820.72 & 39.3 &  300.0& 67.82 $\pm$ 4.22 & -\tabularnewline
2014-06-13 & 56821.69 & 40.3 &  300.7& 67.68 $\pm$ 4.18 &  -\tabularnewline
2014-06-16 & 56824.72 & 43.3 &  302.9& 66.39 $\pm$ 4.16 & -\tabularnewline
2014-06-17 & 56825.74 & 44.4 &  303.6& 65.84 $\pm$ 4.10 & -\tabularnewline
\hline 
\end{tabular}
\end{center}
\end{table}

\subsection{SALT observations}
As discussed in van Soelen et al. (in prep)  optical spectroscopy was undertaken with the South African Large Telescope (SALT) using the Robert Stobie Spectrograph \citep{burgh03}, between 2014 April 30 and 2014 July 21.  The RSS was used in a long slit mode with a wavelength coverage of 6176.6 -- 6983.0~\AA, with a resolution of $R = 11021$ at the central wavelength.   A typical observation consisted of 3-4 camera exposures (total of $\sim 476$ to 500 s) which were co-added to achieve a higher SNR. 

Data were reduced following the standard IRAF procedures, and flux shape correction was done using the spectroscopic standard LTT4364 (observed on 2014 May 11). 


The equivalent width of the H$\alpha$ line was measured by integrating over the line profile (within IRAF) with an assumed linear continuum between the selected positions.  The statistical error in the equivalent width has been estimated using the method discussed in \citet{vollmann06}. 
In order to include an estimate of the uncertainty due to the line continuum, each measurement was performed more than once and any variation in the answer was included in the error estimate.  In general this was smaller than the statistical estimate, though a more rigorous estimate of the error introduced by continuum placement (such as used for the SAAO 1.9-m) will slightly increase the error.  

 Since there was limited overlap between the SALT and SAAO 1.9-m observations (and this is for the SALT data which suffered from instrumental problems), we performed a comparison between the analysis methods used for the different data sets by analysing a selection of spectra from both the SALT and SAAO 1.9m observations using both analysis methods, and
confirmed that both methods gave results that agreed within the uncertainty. The results are listed in Table~\ref{tab:salt}.
\begin{table}
\begin{center}
\caption{Journal of SALT  telescope 2014 observations.}\label{tab:salt} 
\begin{tabular}{|c|c|c|c|c|}
\hline \hline
Date & MJD         &$t-t_{p}$    &$\varphi$& $-W_{H\alpha}$ \\
& (days) & (days) & &(\r{A})\\
\hline
2014-04-01 & 56748.0    & 	 -33.4 &  65.06     &   54.88$\pm$0.68 \\
2014-04-05 & 56752.9  & 	 -28.5 &  70.38     &   55.27$\pm$0.72\\
2014-04-11 & 56758.0    & 	 -23.4 &  77.48     &   55.56$\pm$0.62\\
2014-04-16 & 56763.9  &	 -17.5 &  88.89     &   58.59$\pm$0.10\\
2014-04-25 & 56772.9  &	 -8.5  &  119.94    &   55.37$\pm$0.55\\
2014-04-29 & 56776.8  &	 -4.6  &  143.18    &   56.78$\pm$ 0.69\\
2014-04-30 & 56777.9  &	 -3.5  &  151.17    &   56.55$\pm$ 0.59 \\
2014-05-01 & 56778.8  &	 -2.6  &  158.13    &   57.97$\pm$ 0.54 \\
2014-05-03 & 56780.8  &	 -0.7  &  174.59    &   60.90$\pm$1.02 \\
2014-05-11 & 56788.8  &	 7.4   &  233.86    &   68.46$\pm$ 0.55 \\
2014-05-17 & 56794.8  &	 13.4  &  259.57    &   72.70$\pm$ 0.71 \\
2014-05-23 & 56800.8  &	 19.4  &  275.12    &   71.32$\pm$ 0.71 \\
2014-05-31 & 56808.8  &	 27.4  &  288.14    &   70.62$\pm$2.50 \\
2014-06-06 & 56814.7  &	 33.3  &  294.77    &   68.59 $\pm$4.39 \\
2014-06-12 & 56820.7  &	 39.3  &  299.99    &   64.96$\pm$ 0.70 \\
2014-06-13 & 56821.7  &	 40.3  &  300.75    &   63.70$\pm$ 0.54 \\
2014-06-16 & 56824.7  &	 43.3  &  302.88    &   60.83$\pm$ 0.51 \\
2014-06-17 & 56825.8  &	 44.4  &  303.61    &   61.46$\pm$ 0.75 \\
2014-06-19 & 56827.7  &	 46.3  &  304.82    &   60.54$\pm$ 0.70 \\
2014-06-21 & 56829.7  &	 48.3  &  306.01    &   60.12$\pm$ 0.82 \\
2014-06-22 & 56830.8  &	 49.4  &  306.64    &   60.14$\pm$ 0.65 \\
2014-06-29 & 56837.8  &	 56.3  &  310.21    &   60.15$\pm$3.21 \\
2014-07-09 & 56847.8  &	 66.4  &  314.37    &   57.39$\pm$0.79 \\
2014-07-13 & 56851.7  &	 70.3  &  315.77    &   55.89$\pm$0.49 \\
2014-07-21 & 56859.7  &	 78.3  &  318.33    &   56.69$\pm$ 0.55 \\
\hline
\end{tabular}
\end{center}
\end{table}

\section{X-ray Observations and Results} \label{xr}
We conducted an X-ray monitoring campaign on \psrb\ with \swift\ ,
\nst\ and INTEGRAL telescopes, covering the period between $t_{\rm p} -15$~days and $t_{\rm p}+65$~days. These observations are summarized in Tables~\ref{tab:xrt}, \ref{tab:nustar} and \ref{int_lc}. Tables~\ref{tab:xrt} and \ref{tab:nustar} list identifiers for the data set, UT date, MJD, time relative to periastron passage and exposure time for each observation. In this paper we also use new  \textit{NuSTAR} observations of the nearby X-ray pulsar \rxp\ to reconstruct the broad 0.5 -- 50 keV X-ray spectrum as observed by \szk\ in 2014. All 2014 observations can be well fitted with an absorbed power law model.
 
\subsection{Swift/XRT data reduction}

The {\it Swift} observatory \citep{gehrels04} provides the possibility to
monitor sources of X-ray emission on very different time scales. In
this work we use observations covering almost three months around
periastron passage of PSR B1259-63 between 2014 April 20 and July 8.
In Table \ref{tab:xrt} a log of the used observations is given. The
data were processed using tools and packages available in {\tt
  FTOOLS/HEASOFT 6.14}.

XRT observed PSR B1259-63 both in Photon Counting (PC) and Windowed
Timing (WT) modes. Initial cleaning of events has been done using {\tt
  xrtpipeline} with standard parameters. The further analysis was
performed following \cite{evans2009}. In particular, 
in the PC mode
the source extraction region was a circle with radii from 5 to 30 pixels
depending on the count rate 
\citep{evans2009}; in the WT mode radius of the source
extraction region was 25 pixels. The background was collected over the
annulus region with an inner (outer) radius of 60 (110) pixels in both
observational modes. The count rate from the source was too low to pile up
the detector in all observation except one (00030966025), where we
excluded  the inner region of the source aperture with  a radius of 4
pixels.

The obtained spectra were grouped to have at least 1 count bin$^{-1}$
using the FTOOLS {\tt grppha}. To avoid any problems caused by the
calibration uncertainties at low
energies\footnote{http://www.swift.ac.uk/analysis/xrt/digest\_cal.php},
we restricted our spectral analysis to 1.0 -- 10 keV. The errors
reported in this work are purely statistical and correspond to  a
1$\sigma$ confidence level. 

To estimate the 68\% confidence ranges for the unabsorbed fluxes, we computed the C-statistic likelihood profile on the grid of column density ($n_{H}$) and power law slope ($\Gamma$) values. According to the Wilks theorem, the likelihood values are distributed around the true minimum with a $\chi^2$ distribution with two degrees of freedom, since we have two free parameters. We thus converted the obtained likelihood values to probabilities and performed Monte Carlo simulations in order to convert the computed likelihood profile to the distribution of the unabsorbed fluxes. For each of those distributions (for every single observation) we calculated the mean and its (asymmetric) confidence range by computing the limiting flux values below and above the mean, that contain between them 68\% of all the simulated fluxes.

\begin{table*}
\begin{center}
\caption[Swift/XRT observations of the source PSR B1259-63]{Swift/XRT observations of the source PSR B1259-63} \label{tab:xrt} 
\begin{tabular}{|c|c|c|c|c|c|c|c|}

\hline\hline
Data&Obs Id &  Date & Date,     &  $t-t_P$, & Exposure, & XRT count rate$^{a}$, & XRT\\
Set&       &       & MJD (UTC) & d        & s          & cts s$^{-1}$         &  mode \\
\hline
Sw13 &00030966018 & 2014-04-20 & 56767.1736 & -14.561 & 1171.2 & 0.36 & PC \\
Sw14 &00080099001 & 2014-04-20 & 56767.4857 & -14.249 & 1740.6 & 0.44 & PC \\
Sw15 &00030966019 & 2014-04-27 & 56774.9560 & -6.779 & 4015.7 & 0.29 & PC \\
Sw16 &00030966020 & 2014-05-04 & 56781.9594 & 0.224 & 4075.6 & 0.22 & PC \\
Sw17 &00030966021 & 2014-05-06 & 56783.8933 & 2.158 & 1470.9 & 0.20 & PC \\
Sw18 &00030966022 & 2014-05-07 & 56784.0872 & 2.352 & 2404.9 & 0.19 & PC \\
Sw19 &00030966023 & 2014-05-09 & 56786.1540 & 4.419 & 1475.9 & 0.16 & PC \\
Sw20 &00030966024 & 2014-05-17 & 56794.5981 & 12.863 & 179.8 & 0.44 & PC \\
Sw21 &00030966025 & 2014-05-19 & 56796.8441 & 15.109 & 1987.8 & 0.61 & PC \\
Sw22 &00030966026 & 2014-05-22 & 56799.4340 & 17.699 & 2877.3 & 1.00 & WT \\
Sw23 &00030966027 & 2014-05-25 & 56802.4956 & 20.761 & 2925.1 & 0.74 & WT \\
Sw24 &00030966028 & 2014-05-28 & 56805.6925 & 23.958 & 2991.7 & 0.65 & WT \\
Sw25 &00030966029 & 2014-06-02 & 56810.9245 & 29.189 & 1504.4 & 0.51 & WT \\
Sw26 &00030966030 & 2014-06-03 & 56811.0620 & 29.327 & 970.0 & 0.12 & WT  \\
Sw27 &00030966032 & 2014-06-07 & 56815.0953 & 33.360 & 2696.4 & 0.44 & WT \\
Sw28 &00030966033 & 2014-06-11 & 56819.2927 & 37.558 & 3983.2 & 0.54 & WT \\
Sw29 &00030966035 & 2014-06-12 & 56820.4606 & 38.726 & 999.6 & 0.54 & WT  \\
Sw30 &00030966036 & 2014-06-13 & 56821.8494 & 40.114 & 2036.3 & 0.51 & WT \\
Sw31 &00030966037 & 2014-06-14 & 56822.8488 & 41.114 & 1834.4 & 0.49 & WT \\
Sw32 &00030966038 & 2014-06-15 & 56823.2857 & 41.551 & 767.3 & 0.44 & WT  \\
Sw33 &00030966039 & 2014-06-16 & 56824.5833 & 42.848 & 2026.2 & 0.48 & WT \\
Sw34 &00030966041 & 2014-06-18 & 56826.2590 & 44.524 & 981.5 & 0.50 & WT  \\
Sw35 &00030966043 & 2014-06-25 & 56833.3399 & 51.605 & 3284.0 & 0.28 & PC \\
Sw36 &00030966044 & 2014-06-26 & 56834.5113 & 52.776 & 1513.4 & 0.27 & PC \\
Sw37 &00030966047 & 2014-06-27 & 56835.4127 & 53.678 & 2559.7 & 0.25 & PC \\
Sw38 &00030966048 & 2014-07-02 & 56840.4456 & 58.711 & 2964.3 & 0.26 & PC \\
Sw39 &00030966049 & 2014-07-06 & 56844.7580 & 63.023 & 3264.0 & 0.20 & PC \\
Sw40 &00030966050 & 2014-07-07 & 56845.2536 & 63.519 & 3785.9 & 0.17 & PC \\
Sw41 &00030966051 & 2014-07-08 & 56846.5879 & 64.853 & 3965.7 & 0.18 & PC \\
\hline
\end{tabular}
\end{center}
\vspace{3mm}

\begin{tabular}{ll}

$^{a}$  & Total count rate in 0.5--10 keV energy range \\
\end{tabular}

\end{table*}

\subsection{\textit{NuSTAR} observations and data analysis}

Previous X-ray observations of \psrb\ suffered due to  a lack of high
quality X-ray data above 10 keV. The first hard X-ray measurement of
the system was done with the \textit{OSSE} instrument in  the $40-500$~keV band
during monitoring of the 1994 periastron passage \citep{Grove95}. The {\it OSSE} experiment could not provide spatially-resolved flux
measurements of the source due to its non-imaging design  and, therefore,
flux
pollution from nearby sources could not be fully excluded. The first
attempt to perform hard X-ray imaging of \psrb\ was done with INTEGRAL
 during periastron passage in 2004 \citep{Shaw04},
which demonstrated the importance of spatially-resolved
observations. It was shown that the variable hard X-ray source \rxp\
located  $10^\prime$ away from \psrb, significantly contributes to the hard
X-ray flux measured from this region. Despite  the progress made in understanding the contamination from \rxp\ 
INTEGRAL was unable to provide non-contaminated X-ray measurements of
\psrb\ due to its  $12^\prime$ angular resolution. Later periastron observations
with \szk\ were also not able to 
spatially separate hard X-ray emission from
\psrb\ and \rxp, forcing authors to assume their hard X-ray spectral shapes based
on $2-10$~keV data \citep{Uchiyama09}.  Suzaku observations
indicated the possible break during the first interaction of the
pulsar with the disk.

Significant progress has been recently made thanks to the launch of the
\textit{NuSTAR} mission (Harrison et al., 2013), the first hard X-ray
imaging telescope in space working at energies above 10~keV. \textit{NuSTAR} team initiated
an observational campaign during periastron passage of \psrb\ in 2014. In
total \textit{NuSTAR} performed five observations for a total of
146~ks, targeting the source on-axis of the two co-aligned X-ray
telescopes.  
 The 10 m focusing mirror and detector modules
provide $58^\prime$ (half-power diameter) and $18^\prime$ (FWHM) imaging resolution
over the 3--79~keV X-ray band, with  a spectral resolution of 400 eV
(FWHM) at 10 keV. The {\it NuSTAR} field of view (FOV) is  $12^\prime \times 12^\prime$
at 10~keV as defined by the full width at half intensity. The nominal
reconstructed coordinates are accurate to 8$^{\prime \prime}$ (90\% confidence level)
(Harrison et al. 2013).

We reduced and analyzed data using the \textit{NuSTAR} Data Analysis Software
(NuSTARDAS) v1.3.1, which is 
part of  the HEASOFT 6.15 package.  The data
were filtered for intervals of high background and corrected for
angular offset using catalogued coordinates of \psrb .

\begin{figure*}
\includegraphics[width=\textwidth]{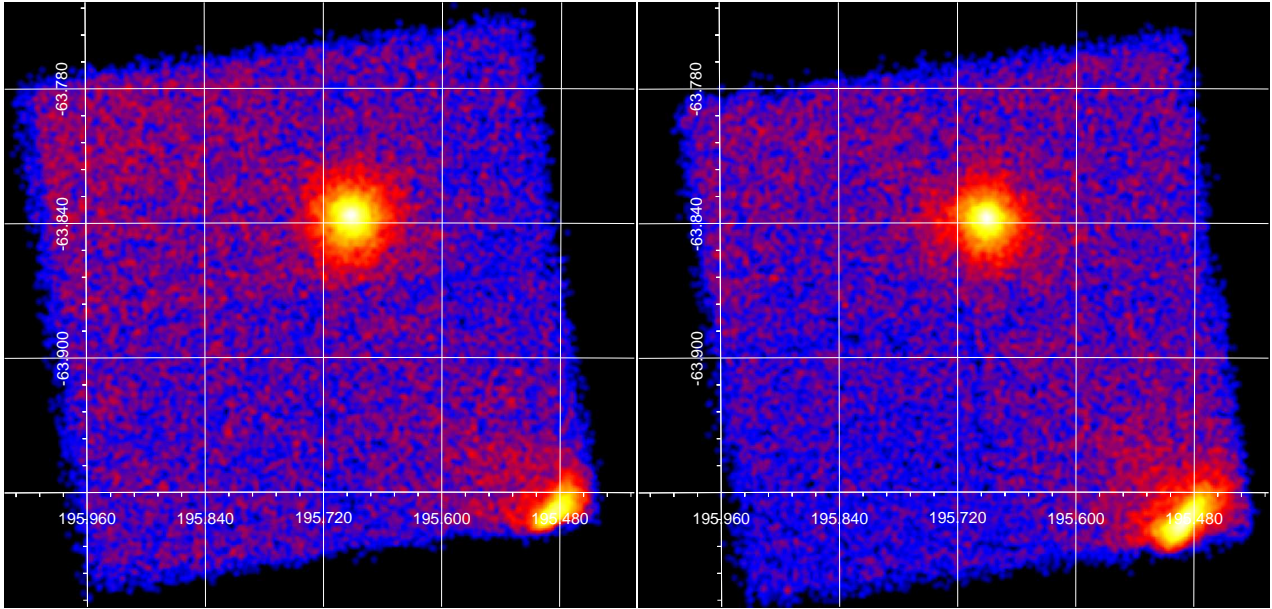}
\caption{{\it NuSTAR} $3-79$~keV image of the region around \psrb\
  (in the centre) obtained with two focal plane modules FPMA
  (left) and FPMB (right). 
   This shows the one out of five observations performed by {\it NuSTAR} where \rxp\ is visible in the lower right corner.
The images are smoothed with $7^{\prime \prime}$ Gaussian kernel for convenience.}
\label{fig:nustar} 
\end{figure*}

During most of the \textit{NuSTAR} observations \rxp\ was out of the
FOV, and only in one of them (ObsID: 30002017004) it
appeared in the corner at  a large off-axis distance, which is
demonstrated in Fig.~\ref{fig:nustar}. It is clearly seen from the
Fig.~\ref{fig:nustar} that \textit{NuSTAR} fully resolved flux from
\psrb\ and \rxp\, allowing one to conduct for the first time hard X-ray
observation of \psrb\ around periastron without any contamination from
the nearby variable and hard \rxp\ source. As discovered by
\cite{Chernyakova_rxp05},  \rxp\ shows coherent spun-up
pulsations at a period of $\sim700$~s. It is interesting to note that
serendipitous {\it NuSTAR} detection of \rxp\ at the corner of the FOV
allowed us to measure its flux pulsations period \citep{krivonos15}, which is in full agreement with
spin-up rate evolution reported in \cite{Chernyakova_rxp05}.

We performed {\it NuSTAR} spectral analysis using {\it nuproducts}
v0.2.5, which is part of NuSTARDAS, to generate {\it NuSTAR} response
matrix (RMF) and effective area (ARF) files for an on-axis point
source. To extract  the source spectrum we utilized circular region centred at the
source position with $70^{\prime \prime}$ radius which comprises $\sim80\%$ of the encircled energy (Harrison et al., 2013). The background spectrum was extracted from  a $115^{\prime \prime}$ circular region positioned away from the source, on the same chip (each detector module has four chips) avoiding chip edges and gaps between them.

\begin{table*}
\begin{center}
\caption[{\it NuSTAR} observations of the source PSR B1259-63]{{\it NuSTAR} observations of the source PSR B1259-63} \label{tab:nustar} 
\begin{tabular}{|c|c|c|c|c|c|c|c|}

\hline\hline
Data&Obs Id &  Date & Date,     & $t-t_p$, & Exposure, & {\it NuSTAR}
count rate$^{a}$, \\
Set&       &       & MJD (UTC) & d        & s          & (FPMA, FPMB) $10^{-1}$ cts s$^{-1}$         \\
\hline
NST1 &30002017002 & 2014-04-20T05:06:07 -- 2014-04-20T22:11:07 &
56767.23 & -14.189&30517.85   &  6.00 , 5.51 \\
NST2 &30002017004 & 2014-05-04T10:01:07 -- 2014-05-05T04:36:07  & 56781.42
& 0.005& 33317.07  &  3.25 , 2.92   \\
NST3 &30002017006 & 2014-05-28T10:11:07 -- 2014-05-28T22:31:07 &
56805.44 & 24.021& 27089.72   & 7.15 , 6.69   \\
NST4 &30002017008 & 2014-06-02T19:21:07 -- 2014-06-03T08:06:07 &
56810.82 &29.405 & 26288.55   & 6.15 , 5.60 \\
NST5 &30002017010 & 2014-06-14T17:21:07 -- 2014-06-15T10:06:07 &
56822.73 &41.312 & 29106.89  &   5.56 , 5.25  \\
\hline
\end{tabular}
\end{center}
\vspace{3mm}

\begin{tabular}{ll}

$^{a}$  & Total count rate in $3-79$~keV energy range \\
\end{tabular}
\end{table*}

\begin{figure}
\includegraphics[width=0.6\columnwidth,angle=270,origin=c]{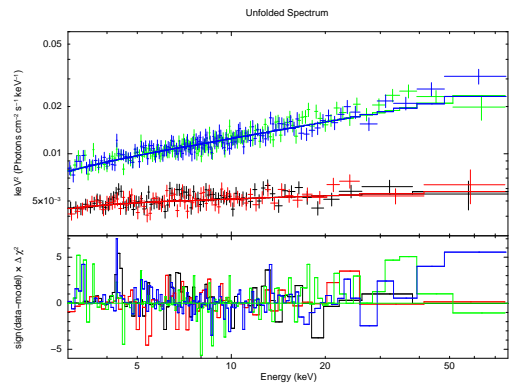}
\caption{Spectra of {\it NuSTAR} NST2 (black and red) and NST3 (green and blue) observations. }
\label{NSt23}
\end{figure}

\subsection{Reanalysis of 2010 \szk\ observations.}
PSR B1259-63 was extensively monitored by \szk\ during its 2007 \citep{Uchiyama09} and 2010 (\cite{Chernyakova_psrb11}, observations Sz9 - Sz11 in Table \ref{Szdata}) periastron passages. \szk\ 2011 broad spectrum of \psrb\ was never presented in the literature. The Suzaku observations were performed with the X-ray
Imaging Spectrometer (XIS; \cite{suzaku_xis}) in 0.3 --
12 keV and the hard X-ray detector (HXD; \cite{suzaku_hxd})
in 13 -- 600 keV. The FOV of both XIS and HXD also contains  the X-ray source, \rxp.  This X-ray pulsar has a luminosity comparable to the one of PSR B1259-63 during the periastron passage and makes significant contribution to the hard
X-ray flux measured by the non-imaging HXD-PIN detector. Recent \textit{NuSTAR} observations of \rxp\ allowed for the first time to measure its broadband spectrum  \citep{krivonos15}. The spectrum of \rxp\  is well described by the absorbed power law with a high energy cut off, $F(\epsilon) = K \epsilon^{-\Gamma} \rm{exp}(-(\epsilon-\epsilon_{c})/\epsilon_{f})$.

\begin{table*}
\begin{center}
\caption{Journal of 2011 \szk\  observations of \psrb. \label{Szdata}}
\begin{tabular}{|c|c|c|c|c|c|c|c|c|c|}
\hline\hline
Data& Date & MJD& $t-t_{\rm p}$&$\phi$&Exposure&$N_{H}$ & $\Gamma$ & F$_{PSRB}$(1-10keV)& F$_{RXP}$(2-10keV)\\
 Set&      & (days)   &  (days)&(deg)&(ks)&     10$^{22}$ cm$^{-2}$& &$10^{-11}$ erg cm$^{-2}$ s$^{-1}$&$10^{-11}$ erg cm$^{-2}$ s$^{-1}$\\
\hline
Sz9 &2011-01-05&55566.8&22&99.6  &90.0&0.54$\pm$0.01 & 1.78$\pm$0.01 &  2.84$\pm$0.01&2.10$\pm$0.03\\
Sz10&2011-01-24&55585.6&41&121.0 &40.3&0.49$\pm$0.01 & 1.54$\pm$0.01 &  1.68$\pm$0.02&2.05$\pm$0.03\\
Sz11&2011-02-02&55594.2&49&126.5 &21.5&0.46$\pm$0.02 & 1.46$\pm$0.02 &  1.47$\pm$0.02&2.26$\pm$0.03\\
\hline
\end{tabular}
\end{center}
\end{table*}

\begin{figure*}
\includegraphics[width=0.36\linewidth]{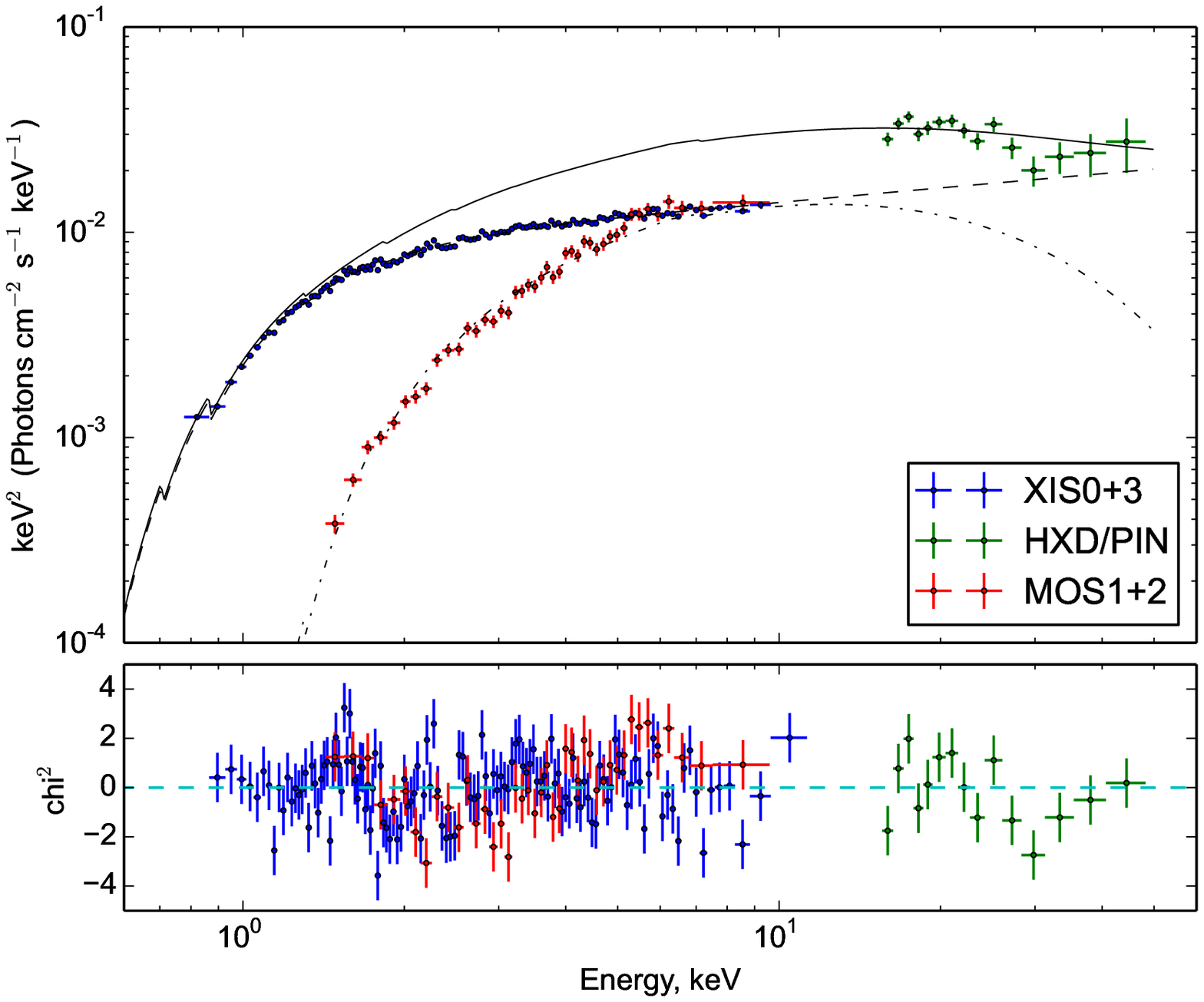}
\hspace{-1.cm}
\includegraphics[width=0.36\linewidth]{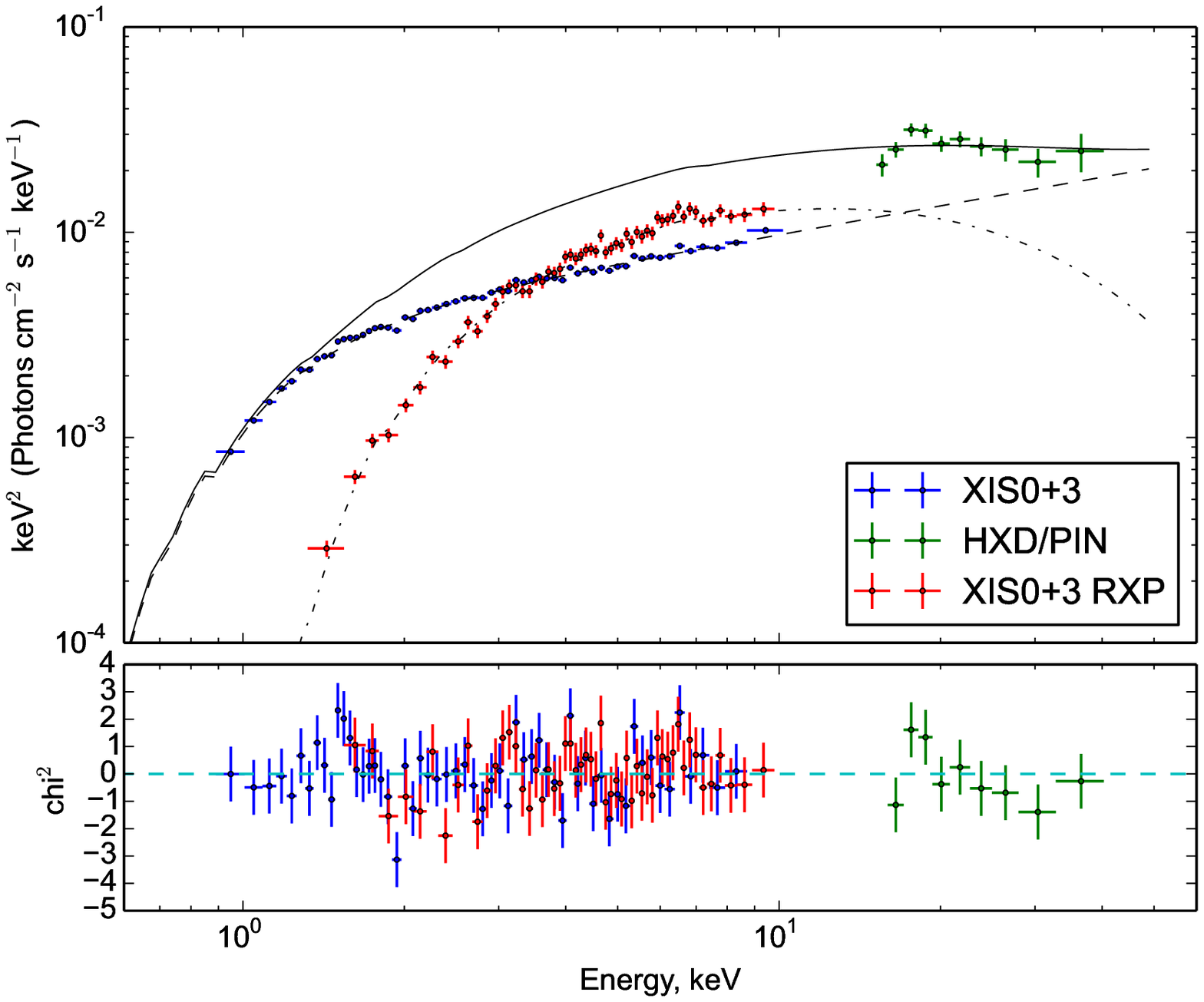}
\hspace{-1.cm}
\includegraphics[width=0.36\linewidth]{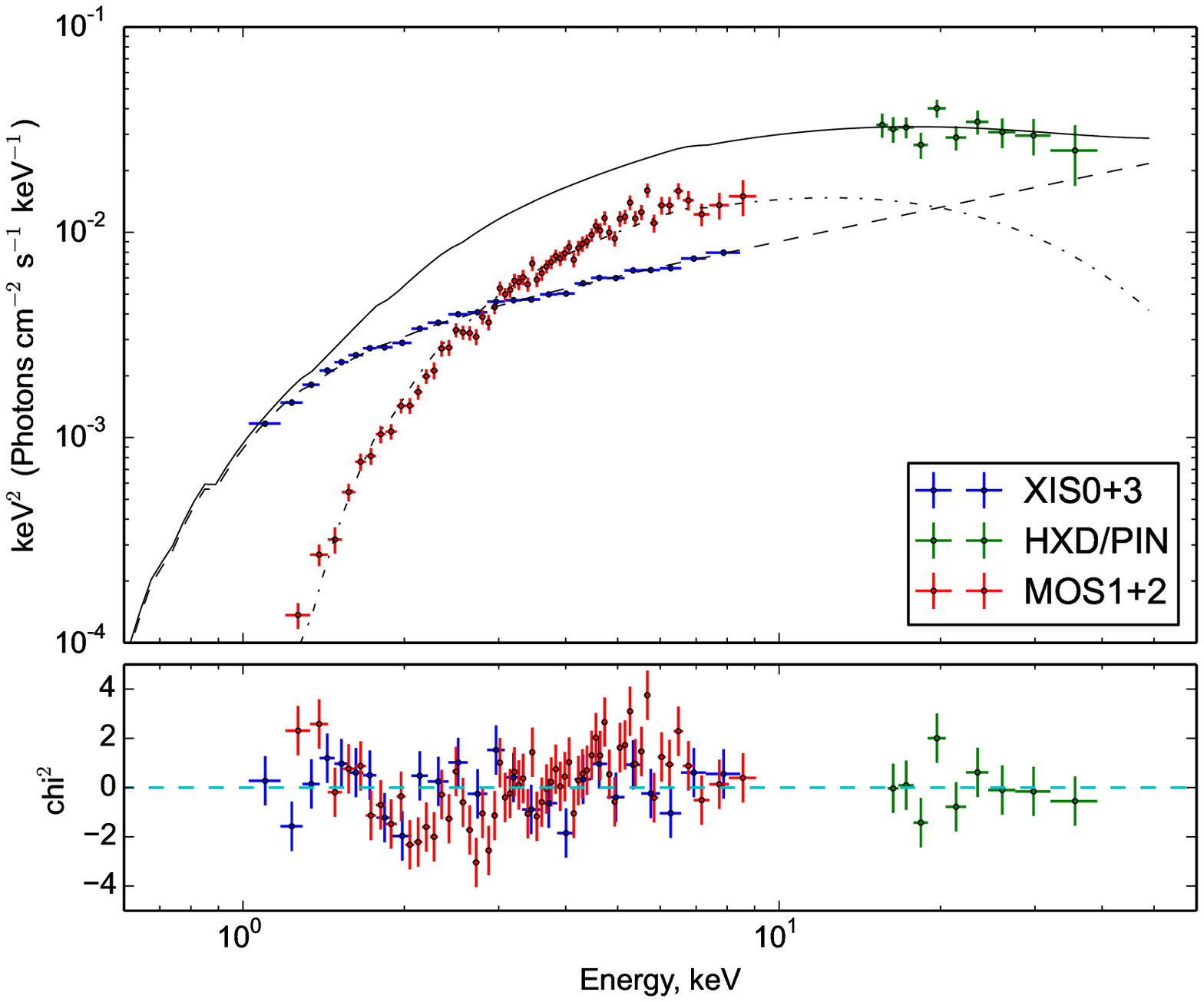}
\caption{ Suzaku XIS (blue crosses) and PIN spectra (green crosses) for S9 (left), S10 (middle) and S11 (rights) observations of 2010. The PIN spectrum  after background subtraction is
decomposed into the two components through joint fitting: PSR B1259-63 power law
model (dashed line), and 2RXP J1301 cutoff power law model (dash-dotted line). \rxp\ was in the field of view only during S10 observation (red crosses). For the reconstruction of \psrb\ PIN spectrum for S9 and S11 observations we have used simultaneous \xmm\ data (red crosses on the  left and right pictures). }
\label{fig:szk}
\end{figure*}
 
To reconstruct the 2011 broadband spectrum of PSR B1259-63 we reanalyzed historical Suzaku data using the HEASOFT software package (version 6.16), with calibration files distributed on 01.07.2014. For the XIS and HXD-PIN, we made use of cleaned event files, in which standard screening was applied. 

HXD data is a sum of emission coming from \psrb\ and \rxp. In our analysis we assumed that the shape of the \rxp\ spectrum was the same in 2010 and 2014, and only normalization changed from one observation to another. The normalization of the \rxp\ spectrum was defined from the simultaneous observations of the imaging instruments, PIN/Suzaku for Sz10, and MOS1,2/\xmm\ for Sz9 and Sz11. These \xmm\ observations are described in \cite{Chernyakova_psrb11}. Simultaneous  fits of Sz9, Sz10, Sz11 and {\it NuSTAR}  data give a good fit ($\chi^2=6795.64$ for 6686 degrees of freedom) with the following parameters: Nh = 2.73$\pm0.07$, $\epsilon_{c} = 6.05\pm0.20$, $\epsilon_{f}= 13.94\pm0.46$, $\Gamma = 1.14\pm0.03$, see Figure \ref{fig:szk}. The  resulting parameters for \psrb\ and 2 -- 10 keV absorbed fluxes for \rxp\ are given in Table \ref{Szdata}.

\subsection{INTEGRAL Observations and data analysis} \label{integral}
INTEGRAL \citep{Winkler03} is a $\gamma$-ray mission covering the energy
the range 15 keV -- 10 MeV. Observations are carried out in individual
Science
Windows (ScW), which have a typical time duration of about 2000s. In 2014
INTEGRAL observed
\psrb\ periastron passage from June 26 to July 10. The data cover
revolutions 1429--1433, adding
up to a total effective exposure time of 141 ks for IBIS/ISGRI in 18-60 keV and 259 ks for JEMX1 \& JEMX2 combined. The data were analyzed using the standard ISDC offline
scientific analysis (OSA) software version 10.0. IBIS/ISGRI images for
each ScW are generated in the 18--60 keV energy band. These images are used to produce the long-term light curve on the
ScW timescale. The total spectrum is obtained using mosaic images as stated
in the IBIS Analysis User Manua\footnote{See http://www.isdc.unige.ch/integral/analysis for more
information.}. To extract the spectra we use the following parameters:
size=1; posmode=-1; widthmode=-1.

\begin{figure}
\includegraphics[width=\columnwidth]{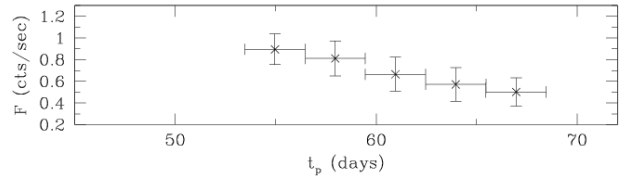}
\caption{\psrb\ 18-60 keV light curve with a 3 day binsize, as observed by IBIS/ISGRI.}
\label{fig:INT_LC}
\end{figure}

PSR B1259-63 was detected with a significance of 3.7 sigma in 3-35 keV
combining JEMX1 \& JEMX2 data. The low significance  hindered further
 investigation of JEMX data. The source was detected at the 11-$\sigma$ confidence level in 18-60 keV energy
range by IBIS/ISGRI. The  resulting  18 -60 keV light curve with a 3 day binsize is
shown in Figure \ref{fig:INT_LC}. The exposure time of each point is given in Table \ref{int_lc}. The light curve shows a graduate decay
of the flux with no striking variability. Please note that some of the flux detected from the position of PSR B1259-63 is actually coming from the nearby source \rxp\ , though as followed from the spectral analysis described below, the total 18-60 keV flux of \rxp\ is  less than a half of the flux of \psrb\ in this energy range.

\begin{table}
\begin{center}
\caption{Integral  observations of \psrb. \label{int_lc}}
\begin{tabular}{|c|c|c|}
\hline \hline
$t-t_{\rm p}$&f(18 -- 60 keV)&Exposure \\
(days) & (cts/sec)&(ks)      \\
\hline
55& 0.90$\pm$ 0.14& 27.21\\
58& 0.81$\pm$ 0.16& 23.65\\
61& 0.67$\pm$ 0.16& 23.27\\
64& 0.57$\pm$ 0.16& 25.42\\
67& 0.50$\pm$ 0.13& 40.27\\
\hline
\end{tabular}
\end{center}
\end{table}

The IBIS/ISGRI angular resolution (12 arcmin)  does not allow us to resolve \psrb\ and  \rxp. Thus similar to the reconstruction of the hard spectrum of Suzaku one needs to fit the INTEGRAL spectrum as a sum of two sources. The quality of the INTEGRAL data  allow us to fit the observed points with a single power law and  the fitting procedure tends to attribute all the observed flux to PSR B1259-63. To prevent this we set a lower limit  for the possible 20 - 60 keV flux  from  \rxp, to $F_{min,RXP}=10^{-11}$~erg~s$^{-1}$~cm$^{-2}$. This minimum value is about a half of the flux registered by  {\it NuSTAR} in May 2015. With such a constraint we get a more realistic fit with $\chi^2=9.93$ for 9 degrees of freedom.
The resulting  20 -- 60 keV flux from \rxp\ varies from $10^{-11}$~erg~s$^{-1}$~cm$^{-2}$ to $1.4\times 10^{-11}$~erg~s$^{-1}$~cm$^{-2}$, and for the PSRB 1259-63 we get the following parameters: F$_{20-60} = 2.13 ^{+0.4} _{-0.5}\times 10^{-11}$~erg~s$^{-1}$~cm$^{-2}$, $\Gamma = 1.67\pm 0.27$. These parameters are in a good agreement with Swift observations taken during the same period (see Figure \ref{fig:gam}).

\subsection{The X-ray light curve}

Panel (b) of Fig.~\ref{fig:MW_LC} shows the X-ray light curve of the system (note that the given 1 -- 10 keV flux is not corrected for the absorption). 
Observations made with different instruments at close orbital phases are
consistent with each other, demonstrating good intercalibration. 

From data obtained at orbital phases similar to the archival observations of
previous periastron passages, one can see that the system orbital light curve is
stable over a time scale of several years.  The stability of the orbital light curve
allows us to use old and new data simultaneously while analysing the orbital
evolution of the flux.  Around the 2014 periastron passage our X-ray observations started  around the time of the first peak  in the X-ray light curve at  $t \sim t_{\rm p}-15$ (Sw13,NSt1).  Similar to previous observations we observed the gradual decrease of the flux, as the pulsar moved towards
the periastron passage. Dense coverage  with \swift\ after the periastron passage allowed us, for the first time, to observe the second rise of the flux in detail, and to  determine that the second X-ray peak is almost twice as high as the first, similar to the behaviour commonly observed in radio. The general stability of the X-ray light curve allows  us to use Chandra 2007 data to conclude that during the second rise the flux doubles in less than three days, and reaches the maximum less than in a week.  After that, the X-ray flux gradually decreased in a good agreement with previous observations. However, detailed \swift\ observations show for the first time that start of GeV flare corresponds to the sharp change of the rate of the X-ray flux decrease, and also to a significant hardening of the X-ray spectrum, as discussed in Section \ref{span}.

\subsubsection{Fast variability in the light curve}
\label{sec:fastvar}

The fast, $\sim 1$~d, variability of the system was previously suggested in \cite{Tam15}. This may indicate that the cooling of electrons in the \psrb\ system happens already  on daily time scales, which challenges the conventional interpretation of the decaying branch of the source light curve in  terms of the gradual cooling of the injected particles. In order to investigate this possibility, we have searched for such fast variability in \swift\ and {\it NuSTAR} 2014 data.

The {\it NuSTAR} data alone, due to their sparse coverage of the source light curve, do not allow  us to look for variability on time scales longer than $\approx 30$~ksec ( the duration of the single {\it NuSTAR} observation). Applying the Structure Function (SF) analysis~\citep{SF_intro,SF_Stephane,savchenko,Me_fast_var} to the light curves, we found no significant variability on  the 0.5--30~ksec time  scale.

Much better coverage of the periastron passage is given by the \swift\ observations,  which provide
light curve coverage in $\approx 10-30$~ksec time bins, separated by $\sim 1$~d time intervals, which creates substantial difficulties for the SF analysis. As a way to overcome this difficulty, \cite{Tam15}~have fitted the light curve with the exponential decay model, demonstrating significant residuals from the fit. However, the intrinsic shape of the decaying branch of \psrb\ light curve may deviate from the simple exponential decay form, e.g. due to the movement of the emitting region through the complex environment of the system and the subsequent change in the cooling rate.

In order to avoid complications due to the uncertain intrinsic light curve shape, we have chosen a different approach. In order to  assess the variability at the time scale $\tau$, we first created an averaged light curve, using $\tau$ as the averaging scale. This averaged light curve contains $n_{av} = (T_{max}-T_{min})/\tau$ time bins. We then compute a difference of the original and the averaged light curves, which then can be quantified in terms of the $\chi^2$ with $n_{orig}-n_{av}$ degrees of freedom. Clearly, a very small value of $\tau$ in this case would result in the averaged light curve perfectly matching the original light curve, resulting in the zero $\chi^2$. At the same time $\tau = T_{max}-T_{min}$ would be equivalent to the fitting the entire light curve to the constant, which in the case of the \psrb\ observations would result in a very large value of $\chi^2$, indicating significant variability.

\begin{figure*}
  \includegraphics[width=\textwidth]{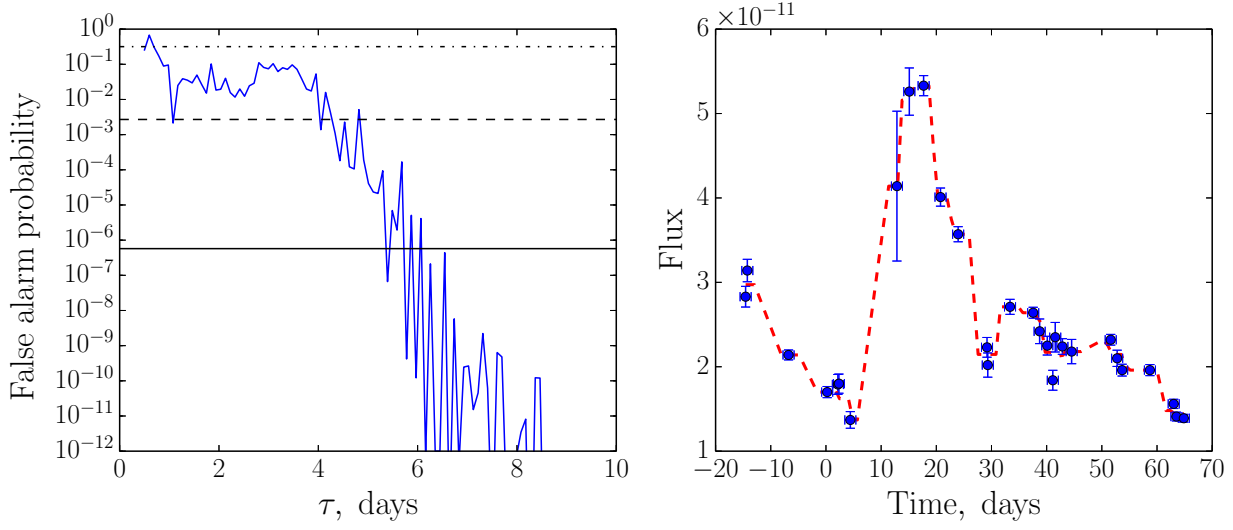}
  \caption{\textit{Left:} false alarm probability of the variability in the \psrb\ light curve from the \swift\ data, see text for details. The 1,3 and 5 $\sigma$ confidence levels are plotted with the dot-dashed, dashed and solid black lines correspondingly. \textit{Right:} The light curve of \psrb\ over the 2014 observational period in the 1-10~keV band, shown together with its averaged version. The averaged light curve is produced with $\tau=4$~days, which corresponds to the minimal scale, where the false alarm probability crosses the $3\sigma$ significance limit.}
  \label{fig:variability_test} 
\end{figure*}

We repeated this procedure for the values of $\tau$ in the range from 0.5 to 10 days with a step of 0.1 day. The result of this scan is shown in Fig.~\ref{fig:variability_test}.
This figure shows that the variability of the source starts to be significant above $\tau \approx 4$~days, corresponding to the rise/decay time scales of the main flares in the light curve. Thus, our analysis disfavours the daily variability of \psrb\, suggesting that its light curve is rather smooth.

\subsection{Spectral Analysis \label{span}}
\begin{figure*}
\resizebox{0.9\hsize}{!}{\includegraphics[angle=0]{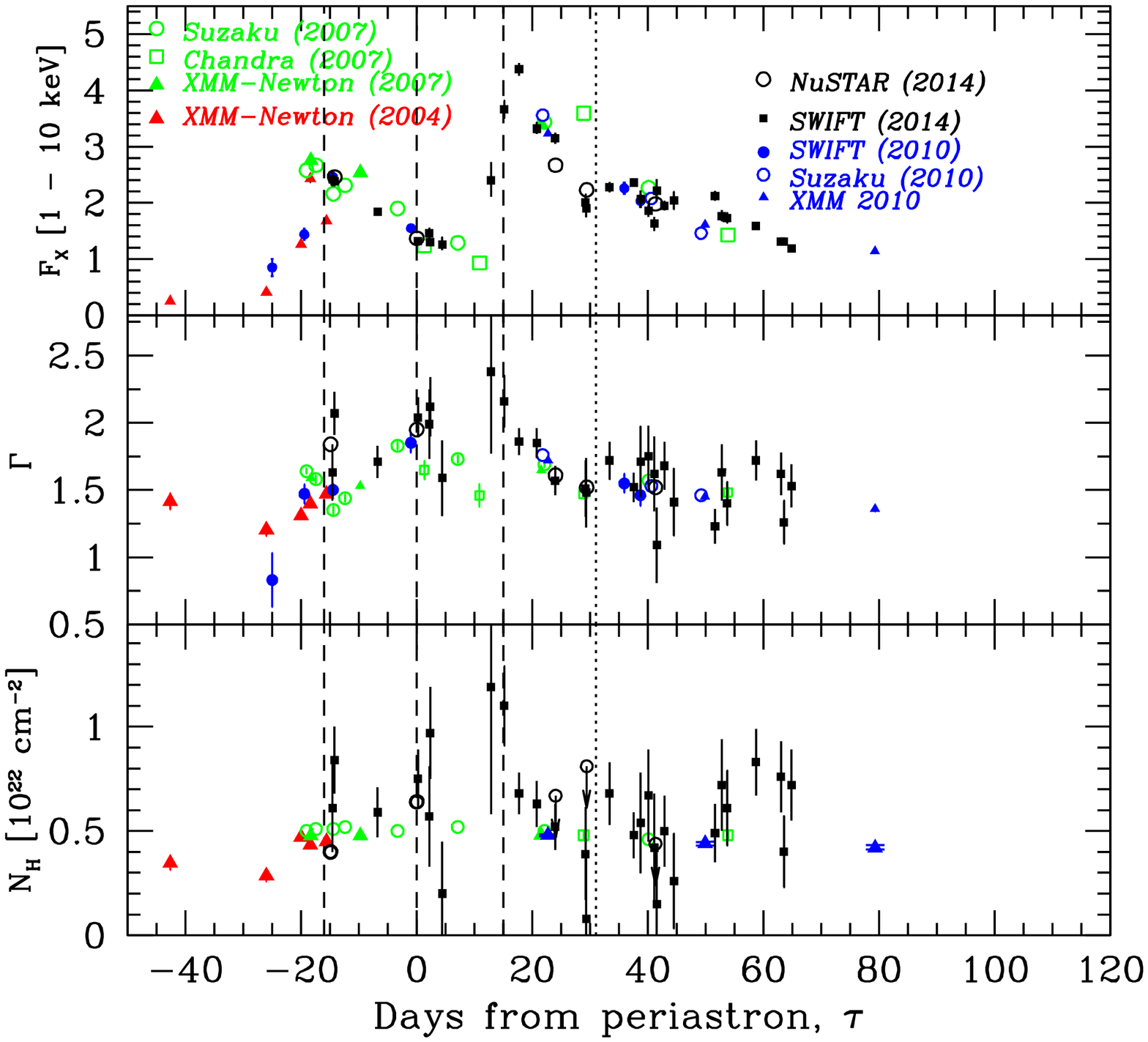}}
\caption{\psrb\ orbital evolution of (1--10 keV) light curve (top panel), spectral index (middle panel) and hydrogen column density (bottom panel), as
seen with \swift\ and \nst\ during the 2014 periastron passage along with the old \xmm\, \szk\ and \swift observations.
1--10 keV flux of the source is given in units of $10^{-11}$~{erg}~{cm}$^{-2}$~s$^{-1}$.}
\label{fig:gam} 
\end{figure*}

The X-ray spectral analyses were done with NASA/GSFC XSPEC v12.7.1 software
package.  A simple power law with a photoelectric absorption describes the data
well, with no evidence for any line features. In Table~\ref{tab:res} we present
the results of the three parameter fit to the \swift\ and \nst\ data. 
The uncertainties are given at the $1\sigma$
statistical level and do not include systematic uncertainties. The graphical representation of the spectral parameters is shown in Figure \ref{fig:gam}. In Figure \ref{fig:gam} new observations are shown along with the ones from previous periastron passages.
The example of the \nst\ spectra (NSt2 and NSt3) are given in Figure \ref{NSt23}.


The average hydrogen column density over the entire Swift 2014 campaign is $n_H=0.588 \pm 0.027$$\times 10^{22}$cm$^{-2}$, estimated as a weighted mean of the values in Table~\ref{tab:res}. A similar analysis of the combined XMM-Newton and Suzaku 2007 data set suggests a lower value of $n_H = 0.487 \pm 0.007$$\times 10^{22}$cm$^{-2}$. The two values are compatible with each other at the $3.6 \sigma$ confidence level. When compared to the averaged value $n_H = 0.493 \pm 0.007$$\times 10^{22}$cm$^{-2}$, the significance of the $n_H$ variation is reduced to $3.2 \sigma$. 

The dotted line in the Figure \ref{fig:gam} shows the moment when the GeV emission was for the first time detected in  the GeV band on a day time scale (31 days after the periastron passage, Caliandro et al. (2015, in press)). It is clearly seen that the rate of the source decay and its spectral index significantly changes after the flare. The decay becomes much more shallow, while the emission becomes much harder, with an average value of $\Gamma=1.5$, instead of $\Gamma=2$ at periastron. No significant change of the column density has been observed.

\begin{table*}
\begin{center}
\caption[Spectral parameters for 2014 observations of PSR B1259-63.]{Spectral parameters for 2014 observations of PSR B1259-63.} \label{tab:res} 
\begin{tabular}{|c|c|c|c|c|c|c|c|}
\hline \hline
Data& $t-t_{\rm p}$,&$F$(1--10~keV)&$F_{abs}$(1--10~keV)&$\Gamma$ &$N_{\rm H}$&C$^*$& ndof\\
Set          &  d&$10^{-11}$ erg cm$^{-2}$ s$^{-1}$ &$10^{-11}$ erg cm$^{-2}$ s$^{-1}$ &&10$^{22}$ cm$^{-2}$& & \\
\hline
Sw13 &-14.56 &  2.83$_{-0.13}^{+0.12}$&2.42$_{-0.13}^{+0.15}$& 1.63$\pm$0.21&0.61$\pm$0.21&189.95&240\\  
Sw14 &-14.24 &  3.14$_{-0.13}^{+0.14}$&2.38$_{-0.09}^{+0.10}$& 2.07$\pm$0.16&0.84$\pm$0.16&260.99&329\\  
NST1 &-14.19 &  2.84$_{-0.05}^{+0.05}$&2.46$_{-0.04}^{+0.05}$& 1.84$\pm$0.02&0.40$\pm$0.26&0.90& 330\\
Sw15 &-6.78  &  2.14$_{-0.06}^{+0.06}$&1.84$_{-0.06}^{+0.07}$& 1.71$\pm$0.12&0.59$\pm$0.12&376.91&402\\ 
Sw16 & 0.22  &  1.70$_{-0.06}^{+0.07}$&1.32$_{-0.05}^{+0.05}$& 2.04$\pm$0.15&0.75$\pm$0.14&295.15&344\\ 
NST2 & 0.005 &  1.88$_{-0.04}^{+0.04}$&1.37$_{-0.03}^{+0.02}$& 1.95$\pm$0.03&0.64$\pm$0.34&0.98&256\\
Sw17 & 2.16  &  1.79$_{-0.12}^{+0.12}$&1.46$_{-0.10}^{+0.10}$& 1.99$\pm$0.26&0.57$\pm$0.24&147.61&190\\  
Sw18 & 2.35  &  1.80$_{-0.12}^{+0.12}$&1.30$_{-0.06}^{+0.07}$& 2.12$\pm$0.22&0.97$\pm$0.22&213.93&247\\ 
Sw19 & 4.42  &  1.37$_{-0.10}^{+0.10}$&1.26$_{-0.10}^{+0.12}$& 1.59$\pm$0.28&0.20$\pm$0.25&120.14&155\\ 
Sw20 & 12.86 &  4.14$_{-0.74}^{+1.04}$&2.40$_{-0.29}^{+0.33}$& 2.38$\pm$0.61&1.19$\pm$0.61&47.14&63\\	
Sw21 & 15.12 &  5.26$_{-0.29}^{+0.27}$&3.66$_{-0.15}^{+0.16}$& 2.16$\pm$0.19&1.10$\pm$0.19&237.50&302\\ 
Sw22 & 17.70 &  5.33$_{-0.12}^{+0.12}$&4.38$_{-0.12}^{+0.11}$& 1.86$\pm$0.10&0.68$\pm$0.10&457.72&529\\ 
Sw23 & 20.76 &  4.01$_{-0.11}^{+0.11}$&3.32$_{-0.10}^{+0.10}$& 1.85$\pm$0.11&0.63$\pm$0.11&403.67&487\\ 
Sw24 & 23.96 &  3.57$_{-0.09}^{+0.09}$&3.15$_{-0.10}^{+0.10}$& 1.57$\pm$0.11&0.52$\pm$0.11&413.16&499\\ 
NST3 & 24.02 &  3.27$_{-0.06}^{+0.03}$&2.67$_{-0.07}^{+0.01}$& 1.61$\pm$0.01&$<0.67$      &      &355\\
Sw25 & 29.19 &  2.23$_{-0.11}^{+0.12}$&2.01$_{-0.12}^{+0.14}$& 1.51$\pm$0.21&0.39$\pm$0.22&214.99&313\\ 
Sw26 & 29.33 &  2.02$_{-0.15}^{+0.14}$&1.90$_{-0.16}^{+0.16}$& 1.48$\pm$0.26&0.08$\pm$0.30&220.49&247\\ 
NST4 & 29.41 &  2.86$_{-0.03}^{+0.03}$&2.23$_{-0.06}^{+0.01}$& 1.52$\pm$0.01&$<0.81$&      &336\\
Sw27 & 33.36 &  2.71$_{-0.09}^{+0.09}$&2.28$_{-0.09}^{+0.09}$& 1.72$\pm$0.14&0.68$\pm$0.15&338.63&430\\ 
Sw28 & 37.56 &  2.64$_{-0.06}^{+0.07}$&2.36$_{-0.08}^{+0.08}$& 1.52$\pm$0.11&0.48$\pm$0.11&355.16&496\\ 
Sw29 & 38.73 &  2.42$_{-0.14}^{+0.15}$&2.07$_{-0.16}^{+0.15}$& 1.71$\pm$0.26&0.54$\pm$0.24&191.01&250\\ 
Sw30 & 40.11 &  2.25$_{-0.11}^{+0.11}$&1.86$_{-0.11}^{+0.11}$& 1.75$\pm$0.23&0.67$\pm$0.22&237.73&319\\ 
Sw31 & 41.11 &  1.84$_{-0.11}^{+0.12}$&1.63$_{-0.13}^{+0.12}$& 1.62$\pm$0.28&0.42$\pm$0.26&201.54&261\\ 
NST5 & 41.31 &  2.61$_{-0.02}^{+0.03}$&1.98$_{-0.05}^{+0.02}$& 1.52$\pm$0.01&$<0.44$      &      &343\\
Sw32 & 41.55 &  2.35$_{-0.18}^{+0.17}$&2.22$_{-0.19}^{+0.20}$& 1.09$\pm$0.28&0.15$\pm$0.29&127.01&203\\ 
Sw33 & 42.85 &  2.24$_{-0.10}^{+0.09}$&1.95$_{-0.09}^{+0.09}$& 1.68$\pm$0.18&0.50$\pm$0.17&281.04&354\\ 
Sw34 & 44.52 &  2.18$_{-0.15}^{+0.14}$&2.04$_{-0.16}^{+0.16}$& 1.41$\pm$0.25&0.26$\pm$0.23&180.19&248\\ 
Sw35 & 51.60 &  2.32$_{-0.07}^{+0.06}$&2.12$_{-0.08}^{+0.09}$& 1.23$\pm$0.13&0.49$\pm$0.14&371.76&418\\ 
Sw36 & 52.78 &  2.10$_{-0.10}^{+0.10}$&1.76$_{-0.09}^{+0.11}$& 1.63$\pm$0.21&0.72$\pm$0.22&210.88&244\\ 
Sw37 & 53.68 &  1.96$_{-0.07}^{+0.07}$&1.73$_{-0.08}^{+0.09}$& 1.40$\pm$0.16&0.61$\pm$0.18&267.29&335\\ 
Sw38 & 58.71 &  1.96$_{-0.06}^{+0.07}$&1.59$_{-0.06}^{+0.06}$& 1.72$\pm$0.15&0.83$\pm$0.16&327.19&358\\ 
Sw39 & 63.02 &  1.56$_{-0.05}^{+0.06}$&1.31$_{-0.06}^{+0.05}$& 1.62$\pm$0.16&0.76$\pm$0.17&262.48&338\\ 
Sw40 & 63.52 &  1.41$_{-0.05}^{+0.05}$&1.31$_{-0.05}^{+0.06}$& 1.26$\pm$0.16&0.40$\pm$0.17&256.06&325\\ 
Sw41 & 64.85 &  1.39$_{-0.04}^{+0.05}$&1.19$_{-0.05}^{+0.05}$& 1.53$\pm$0.16&0.72$\pm$0.17&271.68&343\\ 
\hline
\end{tabular}
\end{center}
$^*$ Relatively poor statistics in some of the Swift spectra forces us to use C statistics in data analysis.
\end{table*}

\section{\emph{Fermi}-LAT Observations and Results} \label{fermi}
 

 The \emph{Fermi}-LAT results included in this paper are adopted from Caliandro et al. (2015, in press). The analysis of \emph{Fermi}-LAT data was performed using the \emph{Fermi} Science Tools\footnote{\url{http://fermi.gsfc.nasa.gov/ssc/}}, 09-34-01 release.
For the 2010 and 2014 periastron passages the analysis was carried out with Pass 7 reprocessed data belonging to the SOURCE event class\footnote{\url{http://fermi.gsfc.nasa.gov/ssc/data/analysis/documentation/Pass7REP_usage.html}}. All gamma-ray photons within an energy range of 0.1--100 GeV and within a circular region of interest (ROI) of 10\deg\ radius centered on \psrb\ were used. To reject  gamma-ray contamination originating 
from the
Earth's limb, we selected events with zenith angle $<$ 100\deg\/. The gamma-ray flux and spectral results of \psrb\ presented in this work were calculated by performing a binned maximum likelihood fit using the Science Tool \emph{gtlike}. The spectral-spatial model constructed to perform the likelihood analysis includes Galactic and isotropic diffuse emission components as well as known gamma-ray sources within 15\deg\ of \psrb\ based on a catalogue internal to the Fermi collaboration ( (now released as 3FGL, \cite{Acero15}). The spectral parameters were fixed to the  catalogue values, except
for the sources within 3\deg\ of \psrb\/\footnote{\url{http://fermi.gsfc.nasa.gov/ssc/data/analysis/scitools/source_models.html}}. For these latter sources, the flux normalization was left free. \psrb\ itself was modelled as a single power-law with all spectral parameters allowed to vary. The contribution of the Galactic and isotropic diffuse emissions within the analyzed ROI was estimated performing a preliminary maximum likelihood fit. For the 2010 periastron event, this fit included data from 5 months prior to 1 year after periastron.  For 2014, this fit included data from 5 months prior to periastron. The resulting scale factors of the isotropic and Galactic diffuse templates were kept fixed in the analysis.

The 2014 \emph{Fermi}-LAT observations showed a persistent nature of the GeV flare with the onset of the flaring activity starting  31 days after periastron (Caliandro et al. (2015, in press)). It turned out that the 2014 GeV flare exhibits a similar average flux level and spectral shape with the flare of 2010. The details of the 2014 GeV flux evolution, differs however from the 2010 light curve, see panel (a) of Figure \ref{fig:MW_LC}. The 2010 flare has a higher peak flux,   and a rapid decreasing evolution (Figure 2a). The 2014 flare has a peak flux lower by a factor of 1.6, which then persists rather than falling rapidly. These differences may be due to inhomogeneities in the shape, density, or extent in the circumstellar disk of the Be star (e.g. the equivalent width of the H$\alpha$ line before the periastron was somewhat higher in 2010, than in 2014, see panel (c) of Figure \ref{fig:MW_LC}).

\section{Discussion } \label{disc}

The broad band non-thermal emission from PSR B1259-63 is produced by high-energy particles accelerated at the shock formed at the interface of the pulsar winds \citep{tavani97}.  The orbital modulation of different components of the broad band spectrum is generally understood as being due to the orbital variability of geometrical parameters, such as the geometry of the contact surface of the stellar and pulsar winds. The overall "two-bump" variability pattern of the source, evident from Fig. \ref{fig:MW_LC}, is generally explained by the passage of the pulsar through the dense equatorial wind of the Be star. This happens  twice 
per orbit, just before and just after the periastron. 

The most puzzling feature of the orbital modulation of the source is the GeV band flaring activity discovered by Fermi/LAT telescope \citep{Abdo2011_b1259,Chernyakova_psrb11}. The flare occurred unexpectedly a month after  periastron 
and ten days after the post-periastron passage of the pulsar through the equatorial wind of the Be star. By this time, the pulsar is already exiting from the dense equatorial stellar wind of the Be star and the overall activity of the source  is decreasing in all energy bands, from radio to X-ray. The sudden brightening of the source at this moment of time was not expected before the Fermi observations. 

The GeV flaring appeared even more puzzling in the absence of clear counterparts to 
the flare in other energy bands. The flare appeared as an "orphan" increase of the GeV flux, not accompanied by
changes in the source state in other wavebands.

The origin of this "orphan" flare was widely discussed in the literature (see e.g. \citet{Petri11,khan12,Kirk13,Dubus13}).  However, no consensus on the nature of the flare was reached, because of the significant lack of information on the flare properties. In particular, before the new  observational campaign around the 2014 periastron passage, 
it was  not known 
if the GeV flare was a unique event or  if it is systematically repeating at each periastron passage. Even if the flare is recurrent from orbit to orbit, it was not clear if it always occurs at the same orbital phase or if it could occur at different phases. It was also unclear if there is only one flare per periastron passage or if the flaring activity could appear as a sequence of flares at different orbital phases. 

New observations in the GeV band in 2014, combined with the multi-wavelength observations reported here provide the missing observational constraints necessary to clarify the nature of the flare. It is now clear that the GeV flaring is recurrent from orbit to orbit. It always occurs at one and the same orbital phase, some 30 days after the periastron passage. 

X-ray observations provide a detailed picture of the behaviour of the source throughout the GeV flaring period and show that the GeV flare is not "orphan". It is accompanied by a specific hard state of the source in the X-ray band (see discussion in Section \ref{span}). 

Optical spectroscopy data reported here also demonstrate that the moment of the onset of the flare is not random. It coincides with a sudden decrease of the equivalent width of H$\alpha$ line, which characterises the state of the equatorial disk of the Be star (see Section~\ref{sec:Opt} and panel (c) of Fig.~\ref{fig:MW_LC}).  

These new observations confirm a model of the flare which was put forward by \cite{Chernyakova_psrb11}. Within this model, the GeV flare occurs at the moment of partial destruction of the equatorial disk of the Be star by the passage of the pulsar. Estimating the mass of the Be star disk from the equivalent width of H$\alpha$ line \citep{Chernyakova_psrb11}, one could find that at the moment of the GeV flare, the disk mass has decreased by a factor of five, from about $2\times 10^{-8}M_\odot$ down to $4\times 10^{-9} M_\odot$, within just five days after the onset of the flare. The nature of such  a dramatic event of destruction of the Be star disk has to be further investigated, but the consequences of the event for the properties of non-thermal emission from the system could be understood in a straightforward way.

Destruction of the equatorial disk also destroys the well-organized geometry of the interacting pulsar / stellar winds system. The regular bow-shaped contact surface of the two winds \citep{tavani97} is destroyed. Instead, a chaotic system of clumps originating from the dense Be star disk 
 produces a contact surface of highly irregular geometry. 

 The destruction of the regular contact surface also closes the escape path for the unshocked pulsar wind which was previously able to escape along a cone  which points 
away from the Be star. The pulsar wind which could not escape releases all its power inside the system. This leads to the increase of the luminosity of the source, up to the $\sim 68\%$ of the spin-down power of the pulsar in 2010 and $\sim 50\%$ in 2014 (\cite{Abdo2011_b1259}, Caliandro et al. (2015, in press)). 

\begin{figure}
\includegraphics[width=\columnwidth]{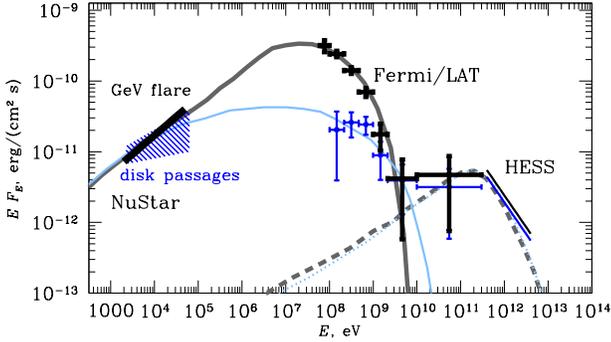}
\caption{Broad band spectrum of the source for the flaring state (black data points) and for the period of periastron passage excluding the GeV flare (blue data points) superimposed on two model spectra from \citet{Abdo2011_b1259} for the synchrotron (solid lines) and inverse Compton (dashed and dotted lines) emission during the GeV flare (grey) and during the rest of the periastron passage (light blue) periods.}
\label{fig:broadband}
\end{figure}

Fig. \ref{fig:broadband} shows a comparison of the new measurements of the spectrum of the flare and pre-flare states of the source with the model of \citet{Abdo2011_b1259}. 
 The Fermi/LAT data from the both observed flares are combined, which allows for a measurement of the \gr\ spectrum down to lower and at higher energies, compared to \citet{Abdo2011_b1259}. Higher quality {\it NuStar} data reveal the X-ray counterpart of the flare. 

These higher quality data are still consistent with the model in which the flare emission is attributed to the synchrotron emission from electrons injected with  a close to monoenergetic spectrum and cooled by the synchrotron energy loss. The synchrotron cooling leads to the formation of a $dN_e/dE\propto E^{-2}$ low energy tail of the electron spectrum. Synchrotron emission from electrons in the tail has the hard spectrum $dN_\gamma/dE\propto E^{-1.5}$  observed in the X-ray band throughout the flare.

Neither the energy nor the origin of the nearly monoenergetic electrons could be firmly established based on the observational data. This is because the maximal energy of the synchrotron emission is close to the self-regulated cut-off in the synchrotron spectrum at $E_{cut}\simeq m_ec^2/\alpha\simeq 100$~MeV, where $\alpha$ is the fine structure constant. Such  a cut-off occurs when electrons are accelerated at  the maximal rate $dE/dt\simeq eB$ and balanced by the synchrotron loss rate. In this regime the maximal energy of the synchrotron emission does not depend on either the magnetic field $B$ or on the energy of electrons. Assuming that the magnetic field in the 
 pulsar and stellar wind interaction zone
is about $B\sim 1$~G, \cite{tavani97} provides an estimate of electron energy $E_e\sim 10^ {14}$~eV, which is necessary to produce synchrotron emission in the 100~MeV energy range. 

Contrary to the flaring component, the persistent component of the broad band spectrum does not have a sharp  peak at 100~MeV. Instead, the X-ray-to-GeV spectrum is flat in the $EF_E$ representation, i.e. it has the form $dN/dE\propto E^{-2}$. Synchrotron emission with such  a spectrum is produced by electrons with a powerlaw spectrum $dN_e/dE\propto E^{-3}$. The synchrotron spectrum hardens below keV  and the slope of the radio-to-X-ray spectrum is $dN/dE\propto E^{-1.5}$, which corresponds  an electron spectrum with  a slope $dN_e/dE\propto E^{-2}$. Thus, the  electron spectrum  
responsible for the persistent emission throughout the two passages of the Be star disk is a broken powerlaw $(dN_e/dE\propto E^{-\Gamma})$  with the slope softening from $\Gamma\simeq 2$ to $\Gamma\simeq 3$ at the energy $E_e\simeq 1\left[B/1\mbox{ G}\right]^{-1/2}$~TeV. The softening by $\Delta\Gamma\simeq 1$ is typical for the effect of the synchrotron cooling on the spectrum. The synchrotron cooling modifies the electron spectrum in the energy range above 1~TeV and it fails to do this below 1~TeV. The absence of cooling in the energy band below 1~TeV could be attributed to the escape of the sub-TeV electrons from the system. The synchrotron cooling time is $t_s\simeq 400\left[B/1\mbox{ G}\right]^{-2}\left[E_e/1\mbox{ TeV}\right]^{-1}$~s. It is comparable to the escape time $t_{esc}=R/s\simeq 300\left[R/10^{13}\mbox{ cm}\right]\mbox{ s}$ for a region which is approximately the size of  
the binary separation distance $R\sim 10^{13}$~cm. 

The injection spectrum of electrons responsible for the persistent emission during the disk passages, $dN_e/dE\propto E^{-2}$ is different from the spectrum of electrons injected during the GeV flare (which is close to mono energetic). This suggests that the two populations of high-energy electrons are produced by different acceleration processes and/or originate from different sources. The natural possibility  for different emission sites is electrons originating from the pulsar wind and electrons from the stellar wind. The two populations of electrons might also appear because of the presence of two different acceleration sites in the system: the non-relativistic shock from the stellar wind side of the contact surface and relativistic shock at the pulsar wind side of the winds contact surface.

In an alternative model of the flaring activity of the source \citep{Abdo2011_b1259}, the "orphan" flare in the GeV band was assumed to be produced via the Bremsstrahlung mechanism. A flare would correspond to a sudden increase of the density of the medium, which could be the debris of the destroyed equatorial disk of the Be star. Such a model is disfavoured by the observations of the X-ray counterpart of the flare. The Bremsstrahlung emission produces  a negligible contribution to the X-ray flux and could not be responsible for the hard spectrum flaring activity coinciding with the GeV flare period. 

Within the synchrotron scenario, the flaring activity of PSR B1259-63 is similar to the flaring activity of the Crab pulsar where the GeV band flares are due to the synchrotron emission from  the highest energy electrons accelerated at maximal possible rate. A study of the comparison of the detailed properties of the GeV flares in the two sources could potentially be useful for understanding the origin of the high-energy electrons and  the location of the high-efficient acceleration sites inside the two sources. Observations reported in this paper show that  the triggering of the flares in PSR B1259-63 occurs simultaneously with an instability which destroys the equatorial disk of the Be star companion to the pulsar. It is not clear if it is the disk instability which triggers an instability of the pulsar wind and of the interface of the pulsar and stellar winds, or the causal link is in the opposite direction. In the case of Crab, the mechanism of triggering the flares  could only be an intrinsic instability of the pulsar wind or of the interface of the pulsar wind and external medium.  An alternative possibility is not available because of the absence of the companion massive star. It remains to be seen if the flares of Crab and PSR B1259-63 are triggered by the same type of instability and are powered by the same type of acceleration mechanism.

\textbf{Acknowledgements. } 
This work was partially supported by the EU COST Action MP1304 "NewCompStar".
The authors thank the International Space Science Institute (ISSI, Bern) for support within the ISSI team ``Study of Gamma-ray Loud Binary Systems'' and SFI/HEA Irish Centre for High-End Computing (ICHEC) for the provision of computational facilities and support. SZh acknowledge support from the Chinese NSFC 11473027, 11133002, XTP project XDA 04060604; Strategic Priority Research Program "The Emergence of Cosmological Structures" of the Chinese Academy of Sciences, Grant No. XDB09000000 and  Strategic Priority Research Program on Space Science, Chinese Academy of Sciences, Grant No.XDA04010300. ST thanks Russian Scientific Foundation for the support (grant 14-12-01287).  MVM is grateful for support from the National Science Foundation through the grant AST-1109247. CK acknowledges the allocation of telescope time by the South
African Astronomical Observatory, and funding from the South
African Research Foundation. The work of IuB was partially supported  by the stipendium of the president of Ukraine (2014-2016). J.L. and D.F.T. acknowledge support from the grants AYA2012-39303, SGR 2014-1073 and support from the National Natural Science Foundation
of China via NSFC-11473027. J.L. acknowledges  support by the Faculty of the
European Space Astronomy Centre.  D.F.T. acknowledges the Chinese Academy of Sciences visiting professorship program 2013T2J0007.

\def\aj{AJ}%
\def\actaa{Acta Astron.}%
\def\araa{ARA\&A}%
\def\apj{ApJ}%
\def\apjl{ApJ}%
\def\apjs{ApJS}%
\def\ao{Appl.~Opt.}%
\def\apss{Ap\&SS}%
\def\aap{A\&A}%
\def\aapr{A\&A~Rev.}%
\def\aaps{A\&AS}%
\def\azh{AZh}%
\def\baas{BAAS}%
\def\bac{Bull. astr. Inst. Czechosl.}%
\def\caa{Chinese Astron. Astrophys.}%
\def\cjaa{Chinese J. Astron. Astrophys.}%
\def\icarus{Icarus}%
\def\jcap{J. Cosmology Astropart. Phys.}%
\def\jrasc{JRASC}%
\def\mnras{MNRAS}%
\def\memras{MmRAS}%
\def\na{New A}%
\def\nar{New A Rev.}%
\def\pasa{PASA}%
\def\pra{Phys.~Rev.~A}%
\def\prb{Phys.~Rev.~B}%
\def\prc{Phys.~Rev.~C}%
\def\prd{Phys.~Rev.~D}%
\def\pre{Phys.~Rev.~E}%
\def\prl{Phys.~Rev.~Lett.}%
\def\pasp{PASP}%
\def\pasj{PASJ}%
\def\qjras{QJRAS}%
\def\rmxaa{Rev. Mexicana Astron. Astrofis.}%
\def\skytel{S\&T}%
\def\solphys{Sol.~Phys.}%
\def\sovast{Soviet~Ast.}%
\def\ssr{Space~Sci.~Rev.}%
\def\zap{ZAp}%
\def\nat{Nature}%
\def\iaucirc{IAU~Circ.}%
\def\aplett{Astrophys.~Lett.}%
\def\apspr{Astrophys.~Space~Phys.~Res.}%
\def\bain{Bull.~Astron.~Inst.~Netherlands}%
\def\fcp{Fund.~Cosmic~Phys.}%
\def\gca{Geochim.~Cosmochim.~Acta}%
\def\grl{Geophys.~Res.~Lett.}%
\def\jcp{J.~Chem.~Phys.}%
\def\jgr{J.~Geophys.~Res.}%
\def\jqsrt{J.~Quant.~Spec.~Radiat.~Transf.}%
\def\memsai{Mem.~Soc.~Astron.~Italiana}%
\def\nphysa{Nucl.~Phys.~A}%
\def\physrep{Phys.~Rep.}%
\def\physscr{Phys.~Scr}%
\def\planss{Planet.~Space~Sci.}%
\def\procspie{Proc.~SPIE}%
\let\astap=\aap
\let\apjlett=\apjl
\let\apjsupp=\apjs
\let\applopt=\ao
\bibliographystyle{mn2e}
\bibliography{Pulsar_Catalog_ALL_Refs_new}

\label{lastpage}
\end{document}